\documentclass[12pt]{article}
\usepackage{amsmath}
\usepackage{amssymb}
\usepackage[mathscr]{eucal}
\usepackage[usenames]{color}
\usepackage[latin5]{inputenc}
\usepackage{url}
\usepackage{longtable}

\newtheorem{thm}{Theorem}

\newtheorem{rmk}{Remark}

\numberwithin{equation}{section}
\numberwithin{thm}{section}
\numberwithin{lemma}{section}
\numberwithin{prop}{section}
\numberwithin{cor}{section}
\numberwithin{rmk}{section}
\numberwithin{defn}{section}

\definecolor{darkolivegreen}{rgb}{0.333333, 0.419608, 0.1843140}

 \DeclareMathOperator{\sn}{sn}

\DeclareMathOperator{\cn}{cn}

\newcommand{\dx}{\partial_x}

\newcommand{\dt}{\partial_t}
\newcommand{\du}{\partial_u}

\begin{document}

\title{\Large Group Classification of a Higher-Order Boussinesq Equation
}

\author{
Y. Hasano\u glu, C. \"{O}zemir\\
\small Department of Mathematics, Faculty of Science and Letters,\\
\small Istanbul Technical University, 34469 Istanbul,
Turkey\thanks{e-mail: ysnhsngl@hotmail.com, ozemir@itu.edu.tr}}

\date{June 29, 2020}

\maketitle

\begin{abstract}
We consider a family of  higher-order Boussinesq equations with an arbitrary nonlinearity. We determine the classes of equations
so that a certain type of Lie symmetry algebra is admitted in this family. In case of a quadratic nonlinearity we provide several exact solutions, some of which are in terms of elliptic functions.
\end{abstract}

\section{Introduction}
The aim of this manuscript is to classify higher-order Boussinesq (HBq) equations  of the form
\begin{equation}\label{hobe}
u_{tt}=\eta_1 u_{xxtt}-\eta_2 u_{xxxxtt}+(f(u))_{xx}
\end{equation}
 according to the Lie symmetry algebras the equation admits depending on the formulation of the nonlinearity $f(u)$ and to  study possible reductions of this equation to find exact solutions. More explicitly, we will determine the classes of functions $f(u)$ for which the equation has finite-dimensional Lie symmetry algebras. Among these classes, we shall concentrate on a specific family, which is widely concerned in literature,  to find exact traveling wave solutions. Here we  assume $\eta_1$, $\eta_2$ are nonzero constants and  $f_{uu}\neq 0$.

The derivation of \eqref{hobe} appears in \cite{rosenau1988dynamics}, when the  approximation of  the equations of motion of  a 1-dimensional lattice to the continuum requires considering higher order effects. It is also derived in  \cite{duruk2009higher} for the propagation of longitudinal waves in an infinite elastic medium within the context of nonlinear non-local elasticity. The authors also investigate the well-posedness of the Cauchy problem.   As a recent literature, we see Eq. \eqref{hobe} in \cite{oruc2019existence} where the authors study the local and global existence and blow-up of solutions to the initial and boundary value problem of the equation. In this literature, $\eta_1$ and $\eta_2$ are positive constants and $f(u)$ is considered to be an arbitrary nonlinearity. HBq equations of  \cite{duruk2009higher} and \cite{oruc2019existence} are obtained from \eqref{hobe} when we replace  $f(u)\rightarrow u+f(u)$.

The Lie symmetry algebra of the Boussinesq equation
\begin{equation}
u_{tt}+uu_{xx}+(u_x)^2+u_{xxxx}=0
\end{equation}
is the Lie algebra of the vector fields
\begin{equation}
D=x\dx+2t\dt-2u \du, \quad P_1=\dx, \quad P_0=\dt,
\end{equation}
which generate translations and dilations, see Refs. \cite{levi1989non,clarkson1989new,nishitani1982similarity,rosenau1986similarity}.
Classical and non-classical similarity reductions of the Boussinesq equation
\begin{equation}
u_{tt}+au_{xx}+b(u^2)_{xx}+cu_{xxxx}=0
\end{equation}
are obtained in \cite{clarkson1989new} and these nonclassical reductions are given a group-theoretical framework in the context of conditional symmetries  in \cite{levi1989non}.

In connection with classification problem in Lie theory, \cite{clarkson1996symmetries} performs the symmetry classification of the generalized Boussinesq equation
\begin{equation}
u_{tt}=u_{xxxx}+(f(u))_{xx}.
\end{equation}
In \cite{gandarias1998classical}, the authors perform Lie symmetry analysis of the equation
\begin{equation}
u_{tt}-u_{xx}+u_{xxxx}+(f(u))_{xx}=0.
\end{equation}
Ref. \cite{bruzon2009travelling} handles the double-dispersion equation
\begin{equation}
u_{tt}=u_{xx}+a u_{xxtt}-bu_{xxxx}+du_{xxt}+(f(u))_{xx}
\end{equation}
and exhibits the functional forms of $f(u)$ so that the equation enjoys Lie symmetry algebras.
Ref. \cite{yu2017lie} studies the symmetry algebra  and reductions of the equation
\begin{equation}
u_{tt}-\Delta u-\Delta u_{tt}+\Delta^2u+k \Delta u_t =\Delta f(u)
 \end{equation}
where $x\in \mathbb{R}^3$ and $f$ is a power-type nonlinearity.  \cite{gandarias2020conservation} considers this equation
for $n=1$, in the form
\begin{equation}
u_{tt}-u_{xx}+au_{xxxx}-bu_{xxtt}=(f(u))_{xx},
 \end{equation}
and for $n=2$, to derive conservation laws. Ref. \cite{recio2016symmetries} considers symmetry algebras of the equation
\begin{equation}\label{B1}
u_{tt}=cu_{xx}+bu_{xxxx}+au_{xxxxxx}+(f(u))_{xx}
\end{equation}
and derives the conservation laws of this equation which admits a Hamiltonian form when written as a system.

Let us finally mention two references which consider the closest family of equations to the one we consider. Ref. \cite{bruzon2009exact} considers
\begin{equation}
u_{tt}=u_{xx}+u_{xxtt}-u_{xxxxtt}-cu_{xxxx}+(f(u))_{xx}
\end{equation}
in the case $c\neq 0$ and finds exact solutions to this equation in terms of trigonometric, hyperbolic and elliptic functions when $f(u)$ has some certain forms. Let us note that the case $c=0$ is not considered in that article separately in the search of the Lie symmetry algebra, therefore they do not cover our results.

Classification of the family of equations \eqref{hobe} in the case $\eta_2=0$; explicitly, the family
 \begin{equation}\label{naeem}
u_{tt}=\delta u_{ttxx}+(f(u))_{xx},
\end{equation}
according to symmetry algebras the equation admits is studied in \cite{ali2015group}. Clearly, our main equation \eqref{hobe} is an extension of this family to the sixth-order. According to the results of \cite{ali2015group}, the Lie symmetry algebra of an equation from the class \eqref{naeem} can be at most three-dimensional. However, we find that for a specific form of $f(u)$, \eqref{hobe} has a four-dimensional symmetry algebra and this result is also valid when $\eta_2=0$, namely, for Eq. \eqref{naeem}. We state this after our main theorem as a Remark, which actually serves as a complementary result to those of \cite{ali2015group}.

Our analysis is consisting of two parts. First we perform the Lie algebra classification of Eq. \eqref{hobe}.  After we  produce some exact solutions for a specific form of $f(u)$.

\section{The Lie Algebra and Reductions }

In what follows we assume that $\eta_1\neq 0$, $\eta_2\neq 0$, $f_{uu}\neq 0$.  The infinitesimal generator is of the form
\begin{equation}\label{}
V=\phi_1(t,x,u)\dt+\phi_2(t,x,u)\dx+\phi_3(t,x,u)\du.
\end{equation}
We find
\begin{equation}
V=\tau(t)\dt+\xi(x)\dx+\phi(t,x,u)\du,
\end{equation}
where
\begin{subequations}\label{det}
\begin{eqnarray}
     &&\phi=Q(x,t)+\big(\frac{1}{2}\tau_t+\frac{3}{2}\xi_x+\phi_0\big)u,\\
     &&\tau_{ttt}=0, \\
     &&2\eta_1\xi_x-5\eta_2\xi_{xxx}=0, \label{14c}\\
     &&8\xi_x-3\eta_1\xi_{xxx}+3\eta_2\xi_{xxxxx}=0,\\
     &&2(\xi_x+\tau_t)f_u+\phi f_{uu}=0,\label{14e}\\
     &&\xi_{xx}f_u+\phi_x f_{uu}=0, \label{14f}\\
     &&Q_{tt}-\eta_1Q_{xxtt}+\eta_2Q_{xxxxtt}-(Q_{xx}+\frac{3}{2}u\xi_{xxx})f_{u}=0,
\end{eqnarray}
\end{subequations}
where $\phi_0$ is an arbitrary constant. If we differentiate \eqref{14e} with respect to $x$ and subtract it from \eqref{14f}, we get $\xi_{xx}=0$ and $\phi_{x}=0$, and hence $Q_x=0$. Eq. \eqref{14c} gives $\xi_x=0$, so $\xi(x)=\xi_0$, a constant. After these, the infinitesimal generator is of the form
\begin{equation}
V=\tau(t)\dt+\xi_0\dx+\phi(t,u)\du
\end{equation}
with
\begin{subequations}\label{det2}
\begin{eqnarray}
     &&\phi=Q(t)+\big(\frac{1}{2}\tau_t+\phi_0\big)u,\\
     &&\tau_{ttt}=0, \quad Q_{tt}=0, \\
     &&2\tau_t f_u+\phi f_{uu}=0. \label{eqf}
\end{eqnarray}
\end{subequations}
 It is seen that when $f$ is arbitrary, we have the two symmetries
\begin{equation}
X_1=\dt, \quad X_2=\dx
\end{equation}
and the symmetry  algebra is the Abelian two-dimensional Lie algebra. One can proceed and solve the system of determining equations  above for different cases. We take another approach and play a little bit on \eqref{eqf} to get an equation involving only $f$. We differentiate \eqref{eqf} with respect to $u$ to get
\begin{equation}\label{eqf2}
2\tau_t f_{uu}+\phi_u f_{uu}+\phi f_{uuu}=0.
\end{equation}
Using \eqref{eqf} and \eqref{eqf2} we can eliminate the term with $\tau_t$ and obtain
\begin{equation}\label{eqf3}
\phi f_u f_{uuu}+\phi_u f_{u} f_{uu}-\phi f_{uu}^2 =0.
\end{equation}
Again we differentiate \eqref{eqf3} with respect to $u$ and find
\begin{equation}\label{eqf4}
2\phi_u f_u f_{uuu}+\phi f_{u} f_{uuuu}-\phi f_{uu} f_{uuu} =0.
\end{equation}
We eliminate $\phi$ between \eqref{eqf3} and \eqref{eqf4} and hence obtain
\begin{equation}\label{eqf5}
f_{u} f_{uu} f_{uuuu}+f_{uu}^2 f_{uuu}-2 f_u f_{uuu}^2=0,
\end{equation}
which is exactly the same equation for $f(u)$ that was obtained in \cite{ali2015group}. Compatible with their findings, Eq. \eqref{eqf5} is solved by the following  different forms of $f$:
\begin{subequations}\label{formf}
\begin{eqnarray}
     &&\mathrm{(a)} \quad f(u)=\alpha e^{\beta u}+\gamma, \label{formfa}\\
     &&\mathrm{(b)} \quad f(u)=\alpha \ln (\beta u + \delta) +\gamma,  \label{formfc}\\
     &&\mathrm{(c)} \quad f(u)=\alpha(\beta u+\delta)^n + \gamma, \quad  n\neq 0,1. \label{formfd}
\end{eqnarray}
\end{subequations}
Here $\alpha, \beta, \gamma, \delta, n $ are arbitrary constants where $\alpha\beta \neq 0 $. Actually, the constant $\gamma$ has no significance when we consider the HBq  equation \eqref{hobe}.   By a transformation $u=\delta_1 \bar{u} + \delta_2$, $\bar{x}=\mu x$, $\bar {t} = \lambda t$ and relabeling the constants,  Eq. \eqref{hobe} with the above forms of $f(u)$ can be converted to an equation with
\begin{subequations}\label{formf2}
\begin{eqnarray}
     &&\mathrm{(A)}\quad f(u)=\alpha e^u, \qquad \alpha=\mp 1, \\
     &&\mathrm{(B)}\quad f(u)=\alpha \ln (u),   \qquad \alpha=\mp 1,\\
    \ &&\mathrm{(C)}\quad f(u)=\alpha u^n,   \qquad \alpha=\mp 1, \quad \mathbb{R}\ni n \neq 0,1. \label{formf2c}
\end{eqnarray}
\end{subequations}
We shall concentrate on these simplified forms of the nonlinearity $f(u)$. Let us note that, in the remaining part of the paper, for all of the cases (A), (B) and (C),  we did not restrict the constant $\alpha$ to $\mp1$ in our calculations, therefore one can use the following results for any nonzero constant $\alpha$.

\vspace{12pt}
\noindent \underline{\textbf{Case A:  $\mathbf{f(u)=\alpha e^u, \quad \alpha = \mp 1}$.}}

\vspace{12pt}
Equation \eqref{hobe} is of the form
\begin{equation}\label{hobea}
u_{tt}=\eta_1 u_{xxtt}-\eta_2 u_{xxxxtt}+\alpha( e^u)_{xx}.
\end{equation}
The Lie algebra  $L_A$ of this equation is three dimensional, $L_A=\{X_1,X_2,X_3\}$, generated by the vector fields
\begin{equation}\label{genA}
X_1=\dt, \quad X_2=t\dt-2 \du, \quad X_3=\dx.
\end{equation}
The nonzero commutation relation is
\begin{equation}\label{com}
[X_1,X_2]=X_1,
\end{equation} therefore the Lie algebra has the structure
$L_A=A_2\oplus A_1=\{X_1,X_2\}\oplus {X_3}$. The optimal system of one-dimensional subalgebras of $A_2\oplus A_1$ is given in \cite{patera1977subalgebras}. Therefore, the optimal system of one-dimensional subalgebras of $L_A$ is
\begin{equation}
\{ X_1\}, \quad \{X_1+\epsilon X_3 \}, \quad \{ -X_2\cos \theta + X_3 \sin \theta\}
\end{equation}
with $\epsilon =\mp 1$ and $0\leq \theta < \pi$.  The reductions through the last subalgebra should be analyzed carefully. \emph{(i)} When $\theta = \frac{\pi}{2}$, we have the generator $X_3=\dx$. Solutions invariant under the group of transformations generated by this subalgebra are time-dependent ones, $u=u(t)$. Not only for \eqref{hobea}, but for any form of $f$ in \eqref{hobe}, these solutions are found from $u_{tt}=0$ hence $u=at+b$, trivially. This subalgebra will not be considered in any of the subcases.  \emph{(ii)} When $\theta \in [0,\pi)-\{\frac{\pi}{2}\}$,  we have  $\{-X_2\cos \theta+X_3 \sin \theta\} \simeq  \{X_2-(\tan \theta) X_3\}$, for  which we simply write $\{X_2+cX_3\}$, $c \in \mathbb{R}$. We observe that when $c=0$, the reduction obtained is 2 less in the order than the order of the reduced equation that is obtained when $c\neq 0$; therefore, for this subalgebra, we consider the cases $c=0$ and $c\neq 0$ separately.

\vspace{12pt}
\noindent \emph{(i) The Subalgebra $X_1=\dt$.} The solutions will have the form $u=u(x)$ and from \eqref{hobea} we get $u=\ln (ax+b)$, where $a,b$ are arbitrary constants.

\vspace{12pt}
\noindent \emph{(ii) The Subalgebra $X_1+\epsilon X_3=\dt+\epsilon \dx$, $\epsilon = \mp 1$.} The invariant solution will have the form $u(x,t)=F(\xi)=F(x-\epsilon t)$. This generator produces traveling wave solutions, and will appear in other forms of the nonlinearity $f(u)$.  Instead of working on \eqref{hobea}, let us do the reduction for \eqref{hobe}, which will be useful for other cases of $f(u)$. (Furthermore, see that since $\dt$ and $\dx$ are symmetries of \eqref{hobe} for any form of $f(u)$, so is the generator $\dt+\epsilon \dx$.) Substituting $u=F(\xi)$, $\xi=x-\epsilon t$ in $\eqref{hobe}$, it  reduces to
\begin{equation}
F''=\eta_1F^{(4)}-\eta_2 F^{(6)}+ [f(F)]''
\end{equation}
which is integrated to
\begin{equation}
\eta_2 F^{(4)}-\eta_1 F''+F-f(F)=K_1 \xi + K_0.
\end{equation}
Here $K_0,K_1$ are arbitrary constants and the derivatives are with respect to the variable $\xi$. Therefore, for $f(u)=\alpha e^u$, the reduced equation is
\begin{equation}
\eta_2 F^{(4)}-\eta_1 F''+F-\alpha e^F=K_1 \xi + K_0.
\end{equation}

\vspace{12pt}
\noindent \emph{(iii) The Subalgebra $X_2=t\dt-2 \du$.} The invariant solution is of the form $u=-2 \ln t+F(x)$, of which substitution into \eqref{hobea} gives $\displaystyle F(x)=\ln (\frac{1}{\alpha}x^2+a x + b)$ and hence
\begin{equation}
u(x,t)=\ln \Big[\frac{1}{t^2}\big(\frac{1}{\alpha}x^2+a x + b\big)\Big].
\end{equation}
\noindent \emph{(iv) The Subalgebra $X_2+cX_3=t\dt+c\dx-2 \du$.} The group-invariant solution will have the form
$u=-2 \ln t+F(\xi)$, $\displaystyle \xi=x-c\ln t. $ From \eqref{hobea} we get, after a further integration,
\begin{equation}
\eta_2c^2 F^{(5)}+c\eta_2F^{(4)}-\eta_1 c^2 F^{(3)}-\eta_1 c F''+c^2F'-\alpha (e^F)'+c F+2\xi=K
\end{equation}
with $K$ being the integration constant.

%
%

\begin{table}[htp]
\centering
\caption{Subalgebras and reduced equations for Cases B and C} \label{table:1}
\begin{tabular}{|l| l | l |}
 \hline
 Subalgebra & $u_{tt}=\eta_1 u_{xxtt}-\eta_2 u_{xxxxtt}+(f(u))_{xx}$ & Similarity variable \\
 \hline\hline

 \emph{\textbf{Case B  }}                 & $f(u)=\alpha \ln u, \quad \alpha=\mp 1$           &\\
 The equation           &  $u_{tt}=\eta_1 u_{xxtt}-\eta_2 u_{xxxxtt}+\alpha(\ln u)_{xx}$&\\
 $L_B=\{X_1,X_2,X_3\}$  & $X_1=\dt, \quad X_2=t\dt+2u\du, \quad X_3=\dx$ &\\
 Reduction by      &       &\\
 $X_1$                  & $u(x)=ae^{bx}$          &\\
 \hline
 $X_1+\epsilon X_3$,
 $\epsilon =\mp 1$      & $\eta_2F^{(4)}-\eta_1 F''+F-\alpha \ln F=K_1\xi+K_0$           &  $u=F(\xi)=F(x-\epsilon t)$\\
 \hline
 $X_2+cX_3$,
 $c\in \mathbb{R}$      & $\eta_2 c^2 F^{(6)} - 3\eta_2 c F^{(5)}+(2\eta_2-\eta_1c^2)F^{(4)}+3\eta_1 cF^{(3)}$ &$u=t^2F(\xi)$ \\
                        &$+(c^2-2 \eta_1)F''-\alpha  (\ln F)''-3cF'+2F=0$      &  $\xi=x-c\ln t$ \\
 \hline
 $X_2$                  &  $\eta_2F^{(4)}-\eta_1 F''-\frac{\alpha}{2} (\ln F)''+F=0$   &  $u=t^2 F(x)$\\
 \hline \hline

 \emph{\textbf{Case C.1}}               & $f(u)=\alpha u^n, \quad \alpha = \mp 1, \quad n\neq 0,1, \quad n \in \mathbb{R}  $         &\\
 The equation           &  $u_{tt}=\eta_1 u_{xxtt}-\eta_2 u_{xxxxtt}+\alpha(u^n)_{xx}$&\\
 $L_{C.1}=\{X_1,X_2,X_3\}$  & $\displaystyle X_1=\dt, \quad X_2=t\dt+\frac{2}{1-n}\,u \,\du, \quad X_3=\dx$ &\\
 Reduction by            &       &\\
 $X_1$                  & $u(x)=(ax+b)^{1/n}$          &\\
 \hline
 $X_1+\epsilon X_3$,
 $\epsilon =\mp 1$      & $\eta_2F^{(4)}-\eta_1 F''+F-\alpha F^n =K_1\xi+K_0$           &  $u=F(\xi)=F(x-\epsilon t)$\\

 \hline
 $X_2+cX_3$,
 $c\in \mathbb{R}$      & $\eta_2 c^2 (n-1)^2 F^{(6)} +\eta_2 c (n-1) (n+3) F^{(5)}$         & \\
                        &$+\big[2\eta_2(n+1)-\eta_1c^2(n-1)^2\big]F^{(4)}$                   & \\
                        &$-\eta_1 c (n-1) (n+3) F^{(3)} +2 (n+1) F$                          &  $u=t^{2/(1-n)}F(\xi)$       \\
                        &$+\Big[c^2 (n-1)^2-2 \eta_1 (n+1)\Big] F''$                         &$\xi=x-c\ln t$     \\
                        &$-\alpha (n-1)^2 (F^{n})''+c(n-1) (n+3) F'=0$                       &     \\
 \hline
 $X_2$                  & $2(n+1)\big(\eta_2F^{(4)}-\eta_1 F''+F\big)-\alpha (n-1)^2 (F^n)''=0$   &  $u=t^{2/(1-n)} F(x)$\\
 \hline \hline

\emph{\textbf{Case C.2}}                 & $f(u)=\alpha u^{-3}, \quad \alpha = \mp 1$                            &\\
 The equation           &  $u_{tt}=\eta_1 u_{xxtt}-\eta_2 u_{xxxxtt}+\alpha(u^{-3})_{xx}$  &\\
 $L_{C.2}$                & $X_1=\dt, \quad X_2=t\dt+\frac{1}{2}u\du, $                  &\\
 $=\{X_1,X_2,X_3,X_4\}$  &  $X_3=t^2\dt+tu \du, \quad X_4 =\dx$                             &  \\
 Reduction by          &                                                                  &\\
 $X_1$                  & $u(x)=(ax+b)^{-1/3}$                                             &\\
 \hline
 $X_4$                  & $u(t)=at+b$                                                      &            \\
 \hline
 $X_2+cX_4$             & $4\eta_2 c^2 F^{(6)} -(\eta_2+4\eta_1 c^2)F^{(4)}$               &  $u=t^{1/2}F(\xi)$      \\
 $(c\geq 0)$                       & $+(\eta_1+4c^2 )F''-4\alpha(F^{-3})''-F=0$                       & $\xi=x-c \ln t$\\
 \hline
 $X_2$                  & $\eta_2F^{(4)}-\eta_1 F''+4\alpha (F^{-3})''+F=0$                &  $u=t^{1/2} F(x)$\\
 \hline
 $-X_1+X_3+dX_4$        & $\eta_2 d^2 F^{(6)}-(\eta_2+\eta_1d^2)F^{(4)}+(\eta_1+d^2)F''$  &  $u=\sqrt{|t^2-1|}F(\xi)$   \\
 $(d\in \mathbb{R})$      & $-\alpha(F^{-3})''-F=0$                                         & $\xi=x+d \tanh^{-1} t$\\
 \hline
 $X_1+\epsilon X_4$,
 $\epsilon =\mp 1$      & $\eta_2F^{(4)}-\eta_1 F''+F-\alpha F^{-3} =K_1\xi+K_0$           &  $u=F(\xi)=F(x-\epsilon t)$\\

 \hline
\end{tabular}
\end{table}

\newpage

\noindent\underline{\textbf{Case B:  $\mathbf{f(u)=\alpha \ln u}$, \quad $\alpha=\mp 1$.}}

\vspace{12pt}
We summarize the results in Table 1. For this case of  $f(u)$, the Lie algebra  $L_B$ of the equation is again three-dimensional, and the basis of the algebra is presented in Table 1. Let us note that the nonzero commutation relation for this algebra is exactly the same as \eqref{com}; therefore, the same Lie algebra is realized as the Case A  by the vector fields that generate the group of transformations of the related equation.

\vspace{12pt}
\noindent\underline{\textbf{Case C:  $\mathbf{f(u)=\alpha u^n}$, \quad $\alpha=\mp 1, \quad \mathbb{R}\ni n \neq 0,1$.}} \; This case has two different branches.

\vspace{12pt}
\noindent\textbf{C.1: $n \neq -3$}. In that case, the Lie symmetry algebra $L_{C.1}$  is 3-dimensional, with the generators given in Table 1.  The structure of the Lie algebra is the same with $L_A$ and  $L_B$. The nonzero commutation relation is as in \eqref{com}.

\vspace{12pt}
\noindent \textbf{C.2: $n=-3$,  $f(u)=\alpha  u^{-3}$}. The symmetry generators of $L_{C.1}$ are also admitted in this case. Besides, there arises a new symmetry generator and the equation admits a 4-dimensional Lie algebra $L_{C.2}=\{X_1,X_2,X_3,X_4\}$,   and the basis of the Lie algebra is presented in Table 1. The nonzero commutation relations are
\begin{equation}
[X_1,X_2]=X_1, \quad [X_1,X_3]=2X_2, \quad [X_2,X_3]=X_3.
\end{equation}

In \cite{patera1977subalgebras} we see this algebra as $A_{3.8}\oplus A_1$ and the optimal system of one-dimensional subalgebras is
\begin{equation}
\{ X_1\}, \quad \{X_4 \}, \quad \{ X_2+cX_4 \}, \quad \{-X_1+X_3+dX_4\}, \quad \{X_1+\epsilon X_4\},
\end{equation}
 where $c \geq 0$ and $d\in \mathbb{R}$. We present the ODEs that are satisfied by the group-invariant solutions of \eqref{hobe} under the transformations generated by these one-dimensional subalgebras in Table 1.

We present the main result of this article in the following Theorem.

\begin{thm}
The Lie symmetry algebra $L$ of the higher order Boussinesq equation \eqref{hobe} can be 2-dimensional, 3-dimensional, or 4-dimensional.
\begin{itemize}
\item[(i)] The Abelian two-dimensional Lie algebra $2A_1$ is admitted as the invariance the algebra of Eq. \eqref{hobe} for any $f(u)$, and is realized by the Lie algebra with basis $\{X_1,X_2\}=\{\dt,\dx\}$.
\item[(ii)] The three-dimensional Lie algebra $A_2\oplus A_1$  (where $A_1$ is the one-dimensional Lie algebra and $A_2$ is the two-dimensional non-Abelian algebra) is admitted as the symmetry algebra of  Eq. \eqref{hobe} if $f(u)$ respects one of the forms given in  \eqref{formf}, or, equivalently, \eqref{formf2}. The related generators of the Lie algebras for these cases are given in, respectively,  for case A in \eqref{genA},  and in case B and case C.1 of Table 1 for the latter two. In all of these cases, the Lie algebra has the decomposition $\{X_1,X_2\}\oplus X_3$.
\item[(iii)] If $f(u)=\alpha (\beta u+\delta)^{-3}+\gamma$, or,equivalently, if $f(u)=\alpha u^{-3}$, $\alpha = \mp 1$,  then $L$ is 4-dimensional, which is denoted as case C.2 in Table 1. The symmetry algebra has the structure
 \begin{equation}\label{decomp}
 L_{C.2} = \{X_1,X_2,X_3\}\oplus X_4 \simeq \mathrm{sl}(2,\mathbb{R})\oplus \mathbb{R}
 \end{equation}
 which contains the simple algebra  $\mathrm{sl}(2,\mathbb{R})$ as a subalgebra.
\item[(iv)] According to these results, maximal dimension of the Lie algebra of a higher-order Boussinesq equation belonging to the class \eqref{hobe} can be 4.
\end{itemize}
\end{thm}

\begin{rmk} Let us have a more close look to the case C.2, i.e., when $f(u)=\alpha u^{-3}$. The Lie algebra with the basis
\begin{equation}
L_{C.2}=\{X_1=\dt, \; X_2=t\dt+\frac{1}{2}u\du, \; X_3=t^2\dt+tu \du, \; X_4 =\dx\}
\end{equation}
is the symmetry algebra of the equation
\begin{equation}
u_{tt}=\eta_1 u_{xxtt}-\eta_2 u_{xxxxtt}+\alpha(u^{-3})_{xx}
\end{equation}
regardless of the values of $\eta_1$, $\eta_2$ and $\alpha$. Therefore, when $\eta_2=0$, the symmetry algebra of the equation
\begin{equation}\label{eta2sifir}
u_{tt}=\eta_1 u_{xxtt}+\alpha(u^{-3})_{xx}
\end{equation}
is also 4-dimensional. Eq. \eqref{eta2sifir} falls into the class \eqref{naeem}, the generalized modified Boussinesq equation, analyzed in \cite{ali2015group}. In their classification of the symmetry algebras of Eq. \eqref{naeem}, they arrive at the same forms of $f(u)$ given in \eqref{formf}, and, according to their results, for these forms of $f(u)$ the symmetry algebras are at most three-dimensional.  As far as we can see, this work does not consider the case $n=-3$, $f(u)=\alpha u^{-3}$ separately and seems to miss the fourth symmetry generator $X_3=t^2\dt+tu \du$ appearing.

Therefore, we should state that, the above Theorem is also valid when $\eta_2=0$, hence the maximal dimension of the symmetry algebra of the generalized modified Boussinesq equation \eqref{naeem}, studied in \cite{ali2015group}, is equal to 4. The simple algebra $L_{C.2}$ with the decomposition \eqref{decomp} is also admitted as an invariance algebra in the case $f(u)=\alpha u^{-3}$.

This remark should be considered  as a complementary result of our main Theorem to the findings in  \cite{ali2015group}.
\end{rmk}

\section{Some Exact Solutions}
In this Section we present some exact solutions to the equation \eqref{hobe}. We consider the nonlinearity $f(u)=\alpha u^2+u$, which gives rise to the equation
\begin{equation}\label{be}
u_{tt}=u_{xx}+\eta_1 u_{xxtt}-\eta_2 u_{xxxxtt}+\alpha(u^2)_{xx}
\end{equation}
where $\alpha\neq 0$ is any constant. The reason for the inclusion of the term $u_{xx}$ is obvious, as one can see from the literature review. For our analysis above, we had considered the $u_{xx}$ term to covered  by the nonlinearity $f(u)$, just on a purpose of bookkeeping. The quadratic nonlinearity $u^2$ can be interpreted like that one considers the stress-strain function of the physical model to be having a quadratic nonlinearity; see \cite{duruk2009higher}.

We aim at finding traveling wave solutions to \eqref{be},  therefore we assume $u=F(\xi)$ with $\xi=kx-ct$ (which amounts to finding the group-invariant solutions under the action of the transformation produced by the generator $c\dx+k\dt$). Putting this ansatz in \eqref{be} and integrating thrice, we obtain
\begin{equation}\label{eqF}
\eta_2 k^4 c^2 \Big[ F'''F'-\frac{1}{2} (F'')^2\Big]-\frac{\eta_1 k^2 c^2}{2} (F')^2-\frac{\alpha k^2}{3}F^3+\frac{c^2-k^2}{2}F^2=K_0.
\end{equation}
We chose the coefficients of the first two integrations as zero and kept only the last one, $K_0$. Since  this equation does not contain the independent variable $\xi$, it can be integrated once by setting $F'=W(F)$ and treating $W$ as the dependent variable and $F$ as independent. However, the resulting equation is so complicated that we could not proceed with it further.

At this point, let us briefly outline the results of \cite{helal2017stability} in their Section 4, in which they consider a $2+1$-dimensional Boussinesq type equation
\begin{equation}\label{2be}
U_{tt}-U_{xx}-U_{yy}-\alpha (U^2)_{xx}-\alpha U_{xxxx}-\alpha \epsilon^2 U_{xxxxxx}=0.
\end{equation}
In order to find traveling wave solutions of this equation, they propose the following ansatz:
\begin{subequations}\label{}
\begin{eqnarray}
     U&=&a_0+a_1\varphi(\zeta)+a_2 \varphi^2(\zeta)+a_3 \varphi^3(\zeta)+a_4 \varphi^4(\zeta), \\
     \zeta(x,y,t)&=&x+y-\kappa\, t,\\
     \Big(\frac{d \varphi}{d\zeta}\Big)^2&=&c_0+c_1\varphi(\zeta)+c_2 \varphi^2(\zeta)+c_3 \varphi^3(\zeta)+c_4 \varphi^4(\zeta). \label{eqfii}
\end{eqnarray}
\end{subequations}
Notice that, if successful, this ansatz will produce trigonometric, hyperbolic or elliptic type solutions due to the Eq. \eqref{eqfii} that $\varphi(\zeta)$ satisfies. It is easy to see that, under the traveling wave ansatz, Eq. \eqref{be} with $u=F(kx-ct)$  and Eq.  \eqref{2be} with $U=U(x+y-\kappa\, t)$ reduce to ordinary differential equations which are the same up to coefficients. Therefore we adapt the methodology in \cite{helal2017stability} find the exact solutions to \eqref{be}. Let us stress that we obtained some more solutions which were not mentioned there.
Therefore, for \eqref{eqF} we propose
\begin{subequations}\label{ansatz}
\begin{eqnarray}
     F(\xi)&=&a_0+a_1\varphi(\xi)+a_2 \varphi^2(\xi)+a_3 \varphi^3(\xi)+a_4 \varphi^4(\xi), \quad \xi(x,t)=k x-ct,\\
     \Big(\frac{d \varphi}{d\xi}\Big)^2&=&c_0+c_1\varphi(\xi)+c_2 \varphi^2(\xi)+c_3 \varphi^3(\xi)+c_4 \varphi^4(\xi)=P(\varphi(\xi)). \label{eqfi}
\end{eqnarray}
\end{subequations}
Upon this substitution, in the resulting expression we express all derivatives of $\varphi(\xi)$ in terms of $\varphi$ using \eqref{eqfi}. Afterwards, we look for the possibility that coefficients of $\varphi^j$, $j=0,1,2,...$ vanish. Below are the several cases we examined.
\subsection{Hyperbolic and trigonometric solutions}

We assume  $c_0=c_1=c_3=0$ and $a_1=a_2=a_3=0$.
We find two main branches for the remaining constants $a_0$, $a_4$, $c_2$, $c_4$, $k$ and $c$.

The first set of parameters is
\begin{subequations}\label{par1}
\begin{eqnarray}
     a_0&=&0,\\
     a_4&=&\frac{840 c^4 c_4^2 (169\eta_2-36\eta_1^2)}{169\alpha},\\
     c_2&=&\frac{13\eta_1}{4c^2 (169\eta_2-36\eta_1^2)},\\
     k^2&=&c^2 \Big(1-\frac{36\eta_1^2}{169\eta_2}\Big)
\end{eqnarray}
\end{subequations}
and the second set of possible parameters is
\begin{subequations}\label{par2}
\begin{eqnarray}
     a_0&=&-\frac{36\eta_1^2}{\alpha (169\eta_2+36\eta_1^2)},\\
     a_4&=&\frac{840 c^4 c_4^2 (169\eta_2+36\eta_1^2)}{169\alpha},\\
     c_2&=&\frac{13\eta_1}{4c^2 (169\eta_2+36\eta_1^2)},\\
     k^2&=&c^2 \Big(1+\frac{36\eta_1^2}{169\eta_2}\Big).
\end{eqnarray}
\end{subequations}
In both cases, $c$ and $c_4$ are arbitrary. Equation \eqref{eqfi} reduces to
$\displaystyle \frac{d \varphi}{|\varphi| \sqrt{c_2+c_4\varphi^2}}=d\xi$, and it is integrated in three different ways depending on the signs of $c_2$ and $c_4$. Observe that $c_2=\eta_1/(52 k^2 \eta_2)$. Although the physical derivation of \eqref{hobe} gives $\eta_1,\eta_2>0$ and   the case $c_2<0$ seems irrelevant,  we include this case also, for completeness.

\vspace{.5cm}
\noindent \underline{\textbf{Case I.a}} \quad  In case $c_2>0$, $c_4>0$, we obtain
\begin{equation}
\varphi(\xi)=\left(\frac{c_2}{c_4}\right)^{1/2}\mathrm{cosech}\Big(\epsilon \sqrt{c_2}(\xi-\xi_0)\Big)
\end{equation}
and the solution to \eqref{be} is
\begin{equation}\label{cosech}
u(x,t)=a_0+\frac{a_4 c_2^2}{c_4^2} \,\mathrm{cosech}^4\Big(\epsilon \sqrt{c_2}(kx-ct-\xi_0)\Big)
\end{equation}
with $\epsilon=\mp 1$. The set of four constants $a_0,a_4,c_2,c_4$ can be chosen as in \eqref{par1} or \eqref{par2}.

\vspace{.5cm}
\noindent \underline{\textbf{Case I.b}}  \quad  If $c_2>0$, $c_4<0$, we obtain
\begin{equation}
\varphi(\xi)=\left(\frac{c_2}{-c_4}\right)^{1/2}\mathrm{sech}\Big(\epsilon \sqrt{c_2}(\xi-\xi_0)\Big)
\end{equation}
and hence
\begin{equation}\label{sech}
u(x,t)=a_0+\frac{a_4 c_2^2}{c_4^2} \,\mathrm{sech}^4\Big(\epsilon \sqrt{c_2}(kx-ct-\xi_0)\Big).
\end{equation}
with $\epsilon=\mp 1$ solves \eqref{be}. When $a_0=0$, this result is   in the same  form with the exact solution presented in \cite{oruc2019existence}.

\vspace{.5cm}
\noindent \underline{\textbf{Case I.c}}\quad  Finally, when  $c_2<0$, $c_4>0$ we find
\begin{equation}\label{sec}
\varphi(\xi)=\left(\frac{-c_2}{c_4}\right)^{1/2}\sec \Big(\epsilon \sqrt{-c_2}(\xi-\xi_0)\Big).
\end{equation}
\eqref{sec} also appears in  \cite{helal2017stability}. The solution to \eqref{be} becomes
\begin{equation}\label{secsol}
u(x,t)=a_0+\frac{a_4 c_2^2}{c_4^2} \sec^4\Big(\epsilon \sqrt{-c_2}(kx-ct-\xi_0)\Big).
\end{equation}
In the solutions \eqref{cosech}, \eqref{sech} and  \eqref{secsol} the valid sets of parameters are those given in \eqref{par1} and \eqref{par2}.
\subsection{Elliptic type solutions}
We assume
$c_4=0$ and $a_1=a_3=a_4=0$.
We find the following values for the remaining constants   $a_0$, $a_2$, $c_0$, $c_1$ and  $c_2$;
\begin{subequations}\label{par3}
	\begin{eqnarray}
	a_0&=&\frac{169\eta_2(c^2-k^2+42c^2k^4c_1c_3\eta_2)-36c^2\eta_1^2}{338k^2\alpha\eta_2},\\
	a_2&=&\frac{105c^2c_3^2k^2\eta_2}{2\alpha},\\
	c_0&=&\frac{4c_1\eta_1}{65k^2c_3\eta_2},\\
	c_1&=&\varepsilon \frac{R}{c_3},\\
	c_2&=&\frac{\eta_1}{13k^2\eta_2},\\
      R&=&\frac{\sqrt{28561(c^2-k^2)^2\eta_2^2-1296c^4\eta_1^4}}{507\sqrt{161}c^2k^4\eta_2^2},
	\end{eqnarray}
\end{subequations}
where  $\varepsilon=\pm{1}$ and $ c_3 $ is arbitrary.\\
Now that we have determined the constants appearing in \eqref{ansatz} successfully, we need to integrate \eqref{eqfi}, which takes the form
\begin{equation}\label{pol2}
\dot \varphi ^2=c_0+c_1\varphi+c_2 \varphi^2+c_3 \varphi^3=P(\varphi),
\end{equation}
and find $\varphi(\xi)$ and hence $u(x,t)$. Evaluation of the integral of  \eqref{pol2} depends on the factorization of the polynomial $P(\varphi)$.
Assume that $\varphi_1$, $\varphi_2$ and $\varphi_3$ are zeros of the equation $P(\varphi)=0$, for which the discriminant is
\begin{equation}
\Delta=18c_0c_1c_2c_3+c_1^2c_2^2-27c_0^2c_3^2-4c_3c_1^3-4c_0c_2^3.
\end{equation}
Making use of \eqref{par3} we obtain
\begin{equation}\label{delt}
\Delta=-\frac{\varepsilon R(80\eta_1^4+7943\varepsilon k^4R\eta_1^2\eta_2^2+2856100k^8R^2\eta_2^4)}{714025c_3^2k^8\eta_2^4}.
\end{equation}
$\Delta=0$ if $R=0$. When we analyze this branch, the coefficients in \eqref{par3} give results the same as in Case I.

Let $\varepsilon=-1$. The sign of $\Delta$ is determined by the sign of the term inside the paranthesis in \eqref{delt}. When we consider this
term as a second-degree polynomial in $R$ and calculate its discriminant, we see it is negative, therefore the polynomial is always positive. Hence $\Delta>0$. Therefore the polynomial \eqref{pol2} has three distinct real zeros $\varphi_1,\varphi_2,\varphi_3$. Then we can factorize \eqref{pol2} as
\begin{equation}\label{pol3}
\dot \varphi ^2=P(\varphi)=c_3 (\varphi-\varphi_1)(\varphi-\varphi_2)(\varphi-\varphi_3).
\end{equation}

\vspace{.5cm}
\noindent \underline{\textbf{Case II.a}} \quad  Let $ c_3>0 $. In order that  \eqref{pol3} makes sense, the right hand side must be nonnegative. Therefore we should consider the  intervals
$\varphi>\varphi_1>\varphi_2>\varphi_3$ and $\varphi_1>\varphi_2>\varphi>\varphi_3$ when integrating  \eqref{pol3}.
Let us first write
\begin{equation}\label{int}
\frac{ d\varphi}{\sqrt{c_3(\varphi-\varphi_1)(\varphi-\varphi_2)(\varphi-\varphi_3)}}=\epsilon d\xi
\end{equation}
where $\epsilon =\mp 1$. In the first hand, when  $\varphi>\varphi_1>\varphi_2>\varphi_3$, using the results available in the handbook \cite{byrd2013handbook}, we  obtain
\begin{equation}
\int_{\varphi_1}^{\varphi}
\frac{d\tau}{\sqrt{c_3(\tau-\varphi_1)(\tau-\varphi_2)(\tau-\varphi_3)}}
=\frac{1}{\sqrt{c_3}}\,g\sn^{-1}\Big(\sqrt{\frac{\varphi-\varphi_1}{\varphi-\varphi_2}},m\Big)
\end{equation}
for the integration of the left hand side of \eqref{int}, where $\displaystyle g=\frac{2}{\sqrt{\varphi_1-\varphi_3}}$, $\displaystyle m^2=\frac{\varphi_2-\varphi_3}{\varphi_1-\varphi_3}$.
This gives rise to  the elliptic function solution  $\varphi$ to \eqref{pol2},
\begin{equation}
\varphi(\xi)=\varphi_1\mathrm{nc}^2\Big(\epsilon\frac{\sqrt{c_3}}{g}(\xi-\xi_0),m\Big)-\varphi_2\mathrm{tn}^2\Big(\epsilon\frac{\sqrt{c_3}}{g}(\xi-\xi_0),m\Big),
\end{equation}
and hence the solution to \eqref{be} can be written as follows
\begin{equation}
u(x,t)=a_0+a_2\left[\varphi_1\mathrm{nc}^2\Big(\epsilon\frac{\sqrt{c_3}}{g}(kx-ct-\xi_0),m\Big)-\varphi_2\mathrm{tn}^2\Big(\epsilon\frac{\sqrt{c_3}}{g}(kx-ct-\xi_0),m\Big)\right]^2.
\end{equation}

\vspace{.5cm}
\noindent \underline{\textbf{Case II.b}}  \quad  When the coefficient $ c_3>0 $, for $\varphi_1>\varphi_2>\varphi>\varphi_3$ we obtain
\begin{equation}
\int_{\varphi_3}^{\varphi}\displaystyle \frac{d \tau}{\sqrt{c_3(\varphi_1-\tau)(\varphi_2-\tau)(\tau-\varphi_3)}}=\frac{1}{\sqrt{c_3}}\,g\sn^{-1}\Big(\sqrt{\frac{\varphi-\varphi_3}{\varphi_2-\varphi_3}},m\Big)
\end{equation}
where $\displaystyle g=\frac{2}{\sqrt{\varphi_1-\varphi_3}}$, $\displaystyle m^2=\frac{\varphi_2-\varphi_3}{\varphi_1-\varphi_3}$. After we find
\begin{equation}
\varphi(\xi)=\varphi_2\mathrm{sn}^2\Big(\epsilon\frac{\sqrt{c_3}}{g}(\xi-\xi_0),m\Big)+\varphi_3\mathrm{cn}^2\Big(\epsilon\frac{\sqrt{c_3}}{g}(\xi-\xi_0),m\Big)
\end{equation}
and hence the solution to \eqref{be} can be written as follows:
\begin{equation}
u(x,t)=a_0+a_2\left[\varphi_2\mathrm{sn}^2\Big(\epsilon\frac{\sqrt{c_3}}{g}(kx-ct-\xi_0),m\Big)+\varphi_3\mathrm{cn}^2\Big(\epsilon\frac{\sqrt{c_3}}{g}(kx-ct-\xi_0),m\Big)\right]^2.
\end{equation}

\vspace{.5cm}
\noindent \underline{\textbf{Case II.c}} \quad  If the coefficient $ c_3<0 $, working on the interval $\varphi_1>\varphi_2>\varphi_3>\varphi$ we  find the  following:
\begin{equation}
\int_{\varphi}^{\varphi_3}\displaystyle \frac{d \tau}{\sqrt{-c_3(\varphi_1-\tau)(\varphi_2-\tau)(\varphi_3-\tau)}}=\frac{1}{\sqrt{-c_3}}\,g\sn^{-1}\Big(\sqrt{\frac{\varphi_3-\varphi}{\varphi_2-\varphi}},m\Big)
\end{equation}
where $\displaystyle g=\frac{2}{\sqrt{\varphi_1-\varphi_3}}$, $\displaystyle m^2=\frac{\varphi_1-\varphi_2}{\varphi_1-\varphi_3}$. This gives us \begin{equation}
\varphi(\xi)=\varphi_3\mathrm{nc}^2\Big(\epsilon\frac{\sqrt{-c_3}}{g}(\xi-\xi_0),m\Big)-\varphi_2\mathrm{tn}^2\Big(\epsilon\frac{\sqrt{-c_3}}{g}(\xi-\xi_0),m\Big)
\end{equation}
therefore the solution to \eqref{be} can be written as follows:
\begin{equation}
u(x,t)=a_0+a_2\left[\varphi_3\mathrm{nc}^2\Big(\epsilon\frac{\sqrt{-c_3}}{g}(kx-ct-\xi_0),m\Big)-\varphi_2\mathrm{tn}^2\Big(\epsilon\frac{\sqrt{-c_3}}{g}(kx-ct-\xi_0),m\Big)\right]^2.
\end{equation}

\vspace{.5cm}
\noindent \underline{\textbf{Case II.d}} \quad  For $ c_3<0 $, on the interval  $\varphi_1>\varphi>\varphi_2>\varphi_3$ we see that we can proceed to obtain
\begin{equation}
\int_{\varphi_2}^{\varphi}\displaystyle \frac{d \tau}{\sqrt{-c_3(\varphi_1-\tau)(\tau-\varphi_2)(\tau-\varphi_3)}}
=\frac{1}{\sqrt{-c_3}}\, g\sn^{-1}\Big(\sqrt{\frac{(\varphi_1-\varphi_3)(\varphi-\varphi_2)}{(\varphi_1-\varphi_2)(\varphi-\varphi_3)}},m\Big)
\end{equation}
where $\displaystyle g=\frac{2}{\sqrt{\varphi_1-\varphi_3}}$, $\displaystyle m^2=\frac{\varphi_1-\varphi_2}{\varphi_1-\varphi_3}$. This immediately results in
\begin{equation}
\varphi(\xi)=\varphi_2\mathrm{nd}^2\Big(\epsilon\frac{\sqrt{-c_3}}{g}(\xi-\xi_0),m\Big)-\varphi_3m^2\mathrm{sd}^2\Big(\epsilon\frac{\sqrt{-c_3}}{g}(\xi-\xi_0),m\Big)
\end{equation}
producing the  solution to \eqref{be}  as
\begin{equation}
u(x,t)=a_0+a_2\left[\varphi_2\mathrm{nd}^2\Big(\epsilon\frac{\sqrt{-c_3}}{g}(kx-ct-\xi_0),m\Big)
      -\varphi_3m^2\mathrm{sd}^2\Big(\epsilon\frac{\sqrt{-c_3}}{g}(kx-ct-\xi_0),m\Big)\right]^2.
\end{equation}

In case $\varepsilon=1$ we have $\Delta<0$. Therefore the polynomial \eqref{pol2} has one real zero $\varphi_1$ and two complex conjugate zeros $\varphi_2,\varphi_3$.

\vspace{.5cm}
\noindent \underline{\textbf{Case II.e}}  \quad If the coefficient $ c_3>0 $ we can obtain
 \begin{equation}
 \int_{\varphi_1}^{\varphi}\displaystyle \frac{d \tau}{\sqrt{c_3(\tau-\varphi_1)[(\tau-b_1)^2+a_1^2]}}=\frac{1}{\sqrt{c_3}}\, g\cn^{-1}\Big(\frac{A+\varphi_1-\varphi}{A-\varphi_1+\varphi},m\Big)
 \end{equation}
where $\displaystyle b_1=\frac{\varphi_2+\varphi_3}{2}$,  $\displaystyle a_1^2=-\frac{(\varphi_2-\varphi_3)^2}{4}$, $A^2=(b_1-\varphi_1)^2+a_1^2$, $\displaystyle g=\frac{1}{\sqrt{A}}$,    $\displaystyle m^2=\frac{A+b_1-\varphi_1}{2A}$. After that  we  find
\begin{equation}
\varphi(\xi)=\varphi_1+A\frac{1-\cn\big(\epsilon\frac{\sqrt{c_3}}{g}(\xi-\xi_0),m\big)}{1+\cn\big(\epsilon\frac{\sqrt{c_3}}{g}(\xi-\xi_0),m\big)}
\end{equation}
and hence the solution to \eqref{be} can be written as
\begin{equation}
u(x,t)=a_0+a_2\left[\varphi_1+A\frac{1-\cn\big(\epsilon\frac{\sqrt{c_3}}{g}(kx-ct-\xi_0),m\big)}{1+\cn\big(\epsilon\frac{\sqrt{c_3}}{g}(kx-ct-\xi_0),m\big)}\right]^2.
\end{equation}

\vspace{.5cm}
\noindent \underline{\textbf{Case II.f}} \quad If the coefficient $ c_3<0 $ one can proceed to get
\begin{equation}
\int_{\varphi}^{\varphi_1}\displaystyle \frac{d \tau}{\sqrt{-c_3(\varphi_1-\tau)[(\tau-b_1)^2+a_1^2]}}=\frac{1}{\sqrt{-c_3}}\, g\cn^{-1}\Big(\frac{A-\varphi_1+\varphi}{A+\varphi_1-\varphi},m\Big)
\end{equation}
where $\displaystyle b_1=\frac{\varphi_2+\varphi_3}{2}$, $\displaystyle a_1^2=-\frac{(\varphi_2-\varphi_3)^2}{4}$,  $\displaystyle A^2=(b_1-\varphi_1)^2+a_1^2$, $\displaystyle g=\frac{1}{\sqrt{A}}$,  $\displaystyle m^2=\frac{A-b_1+\varphi_1}{2A}$.\\
After this we get
\begin{equation}
\varphi(\xi)=\varphi_1-A\frac{1-\cn\big(\epsilon\frac{\sqrt{-c_3}}{g}(\xi-\xi_0),m\big)}{1+\cn\big(\epsilon\frac{\sqrt{-c_3}}{g}(\xi-\xi_0),m\big)}
\end{equation}
and hence the solution to \eqref{be} turns out to be
\begin{equation}
u(x,t)=a_0+a_2\left[\varphi_1-A\frac{1-\cn\Big(\epsilon\frac{\sqrt{-c_3}}{g}(kx-ct-\xi_0),m\Big)}{1+\cn\Big(\epsilon\frac{\sqrt{-c_3}}{g}(kx-ct-\xi_0),m\Big)}\right]^2.
\end{equation}

\section{Conclusion}
In this work, we  considered a higher-order Boussinesq equation with an arbitrary nonlinearity $f(u)$. We determined the canonical forms of $f(u)$ so that the equation admits certain finite-dimensional Lie algebras. We proved that, within this family, the maximal dimension of the Lie algebra of the equation is equal to 4, and this is realized when $f(u)$ assumes some definite form. This result is also true in the case of a generalized modified Boussinesq equation, when $\eta_2=0$.

After that, we considered the case where $f(u)$ is a second degree polynomial in $u$. We produced some exact solutions which were expressed in terms of trigonometric, hyperbolic and elliptic functions. To our knowledge, among the nine solutions we were able to find, except the one in \eqref{sech}, all the other eight given in Case I and Case II appear in literature the first time for the higher-order Boussinesq equation with quadratic nonlinearity.

In this manuscript, we restricted ourselves to a subclass of \eqref{formfd}, by searching for the exact solution in the case  $f(u)=u+\alpha u^2$. Actually, the analysis of the reduced equation for $n=2$ for Case C.1 in Table 1 which were obtained by the infinitesimal generator $X_1+\epsilon X_3$ would follow similar lines to the analysis in Section 3. Mainly due to the complicated nature of the reduced equations, we do not perform a further analysis for the reduced equations in this work. Regarding the families \eqref{formf} or equivalently \eqref{formf2},  which mean some HBq equations with certain symmetries, one can ask another question: Do these canonical forms of nonlinearities $f(u)$ have any physical meaning? As far as we know, the answer is affirmative when $f$ has power-type nonlinearities like $f(u)=u+\alpha u^2$, etc., and that has been the main reason for writing Section 3 of this manuscript. The analysis of the other reduced equations remains still open.

\bibliographystyle{unsrt}

\end{document}